\documentclass[aps,prl,twocolumn,footinbib,superscriptaddress,longbibliography,amsmath,amssymb]{revtex4-2} 

\usepackage{graphicx}
\usepackage{subfigure}
\usepackage{epsfig} 
\usepackage{dcolumn}
\usepackage{bm}
\usepackage{times} 
\usepackage{amsmath}
\usepackage{float}
\usepackage{color}
\usepackage{multirow}
\usepackage{hyperref}
\usepackage[normalem]{ulem}
\usepackage{xcolor}
\definecolor{darkblue}{rgb}{0,0,0.7}
\hypersetup{
    colorlinks=true,
    linkcolor=darkblue,     
    citecolor=darkblue,     
    filecolor=magenta,      
    urlcolor=darkblue       
}

\begin{document}
\title{The Andreev-Ising-Josephson Diode}

\author{Sourabh Patil}
\affiliation{Fachbereich Physik, Universit\"{a}t Konstanz, D-78457 Konstanz, Germany}
\author{Gaomin Tang}
\email{gmtang@gscaep.ac.cn}
\affiliation{Graduate School of China Academy of Engineering Physics, Beijing 100193, China}
\author{Wolfgang Belzig}
\email{wolfgang.belzig@uni-konstanz.de}
\affiliation{Fachbereich Physik, Universit\"{a}t Konstanz, D-78457 Konstanz, Germany}

\begin{abstract}
The transition-metal dichalcogenides featuring Ising spin-orbit coupling in so-called Ising superconductors offer a unique system to study the interplay of singlet and triplet superconductivity. 
The presence of high critical fields, spectral properties such as the mirage gap, and field-tunable charge and spin currents in Ising-superconductor Josephson junctions are some of the important features.
In this work, we study Ising-superconductor Josephson junction with a transparent interface and show that Andreev bound states are spin-split due to a relative misorientation of in-plane fields in the superconducting contacts. Correspondingly, supercurrent-phase relations display a strongly non-sinusoidal behavior.
Introducing additional spin-polarized channels with low transmission results in a nonreciprocal current-phase relation with a diode effect that can be tuned by the in-plane exchange fields. The diode efficiency reaches high values of the order of 40\% and is not sensitive to disorder in the junction. Such structures can be realized in van-der-Waals heterostructures of two dimensional superconductors and magnets.
\end{abstract}

\maketitle
{\it Introduction.--}
The interplay of superconductivity and magnetism has recently given rise to exciting physics, contributing to the field of superconducting spintronics.  
Transition-metal-dichalcogenides which consist of monolayer to few-layer
superconductors with an intrinsic Ising spin-orbit coupling (ISOC) have emerged
as suitable platform in this regard~\cite{Lu15, Saito16, Xi16, Xing17, Dvir18,
Costanzo18, Lu18, delaBarrera18, Sohn18, Li20, Cho21, Hamill21, Idzuchi21, Ai21,
Kuz21}. 
The ISOC makes these superconductors robust against in-plane magnetic fields~\cite{Bulaevskii76, Gorkov01, Frigeri04} much higher than the Pauli limit~\cite{Chandrasekhar, Clogston}.
Therefore, superconductivity in the presence of strong ferromagnetism can be studied in such Ising superconductors. The in-plane field in these materials induces spin-triplet correlations~\cite{Rahimi17,
Moeckli18, Moeckli20, Moeckli20_DOS, mirage, Kuz21}, leading to the prediction of spectral features such as mirage gaps~\cite{mirage, ising_patil, ilic_mirage_2023}.
Ferromagnetic (FM) Josephson junctions have been predicted to show $0$ and $\pi$ states, useful to construct $\phi_0$-junctions~\cite{Idzuchi21,Ai21}. 
FM Ising-superconductor junctions~\cite{Zhou16,Transport_Sun18} and Josephson junctions with half-metal barriers~\cite{Transport_Sun19} focus on spin-triplet Andreev reflection. 
The charge and spin currents in Ising-superconductor Josephson junctions are shown to be field-tunable~\cite{GT21}.

Andreev bound states (ABSs) exist in the proximity of superconductors, often localized within the region of superconducting barriers, enabling the flow of supercurrent~\cite{Beenakker_1991, Furusaki_1991}.
The ABSs play a crucial role in superconducting hybrid structures opening the possibilities for topologically protected qubits for quantum computation~\cite{nazarov-2003,Zazunov_2003,nazarov_2010}.
However, the study of ABSs in the Ising-superconductor Josephson junctions has received limited attention.

Moreover, recent advancements in the field of diode effect for superconductors have gathered significant attention, in the pursuit of nonreciprocal current transport~\cite{diode_grein,Yokoyama_Eto_Nazarov_2014,Dolcini_Houzet_Meyer_2015,diode_ando, fominov_JDE, diode_Souto, diode_intrinsic, diode_theory_rashba, diode_general-theory, diode_field-free, diode_spintronic, diode_JJ,diode_phenomenological, diode_finite-mom, diode_universal, diode_JDE4}. 
The breaking of inversion or time-reversal symmetry is necessary for an asymmetric current-phase relation (CPR) as a function of the Josephson phase~\cite{diode_general-theory}. 
A supercurrent interferometer with two junctions with one non-sinusoidal CPR and one sinusodial CPR
is the most basic textbook example for a diode effect~\cite{fominov_JDE,diode_Souto}. Note that this proposal requires to flux bias the SQUID for the diode effect to appear.

\begin{figure}[thb]
    \centering
        \includegraphics[width= 0.9\columnwidth]{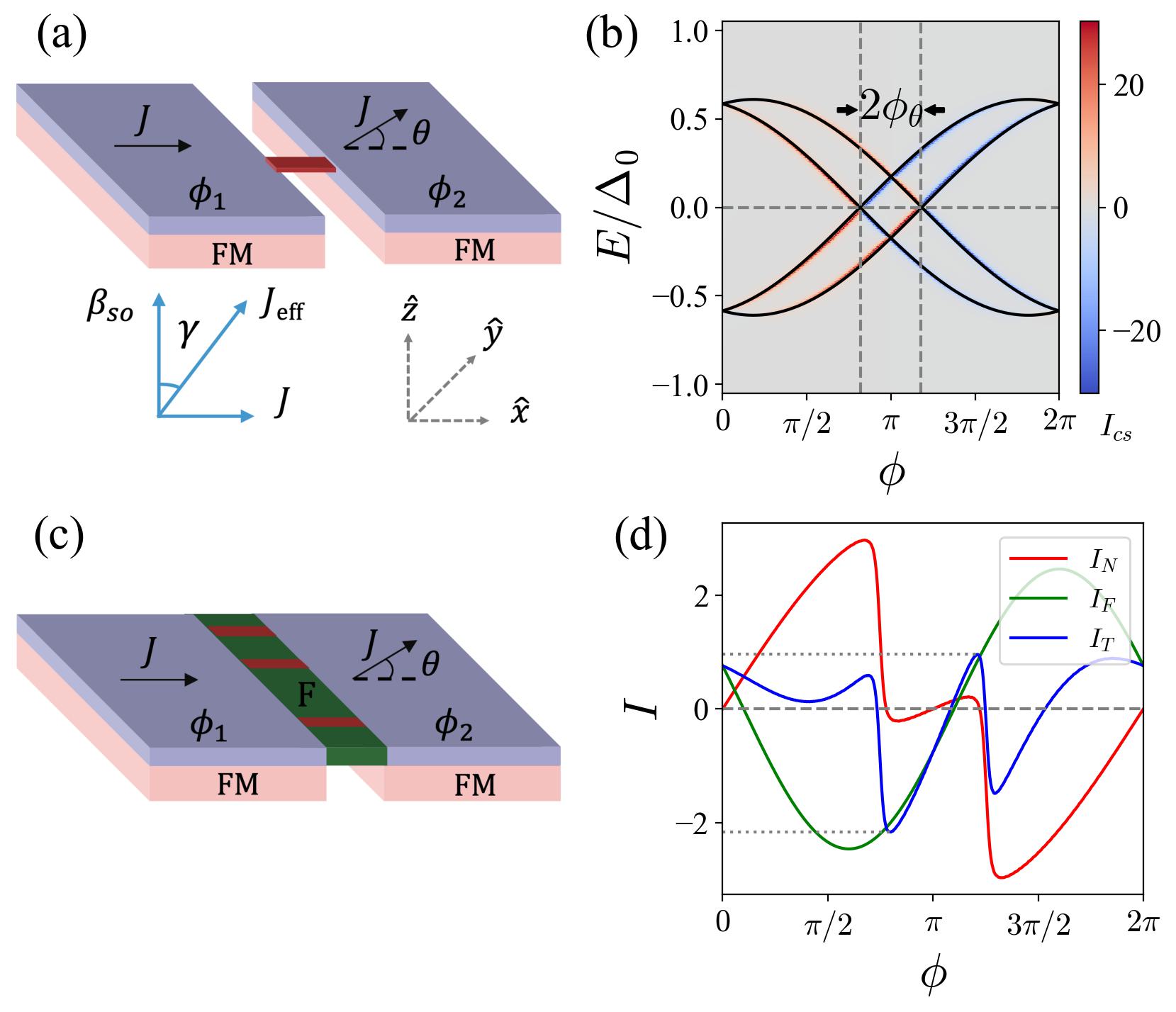}
    \caption{(a) Schematic plot of the Ising-superconductor Josephson junction with a nonmagnetic channel of high transmission. The relative angle between the exchange fields is $\theta$ and the phase difference is $\phi = \phi_1 - \phi_2$. The effective exchange field $\bm{J}_{\rm eff}$ has an angle $\gamma$ to spin-orbit coupling.
    (b) Andreev bound states (black) and spectral supercurrent (color scale) for $\theta = \pi/2$, $\beta_{so}=7\Delta_0$, and $J =3\Delta_0$ (black curve). The spin splitting is parameterized by $\phi_\theta$ illustrating the spin-splitting by the non-collinear in-plane exchange fields.
    (c) Schematic plot of the Andreev-Ising-Josephson junction. A spin-polarized magnetic barrier (in green) and nonmagnetic transmissive channels (in red) connect two superconductors.
    (d) Current-phase relations for the junctions with a transparent nonmagnetic barrier (red curve) and a ferromagnetic tunneling barrier (green curve). The total current-phase relation (blue) is highly nonreciprocal with positive and negative critical currents $I_c$ indicated by dotted lines. The conductance ratio of the nonmagnetic and the magnetic parts is 1:10, while $J= 3\Delta_0$, $T=0.01 T_c$, and $\theta= 0.9\pi$ and supercurrents are in the units of $2G_0 \Delta_0 /e$. }
    \label{figure1}
\end{figure}

Motivated by these developments, we explore the ABS spectrum in a single-channel Ising-superconductor Josephson junction with a transparent nonmagnetic constriction as shown in Fig.~\ref{figure1}(a). 
The FM layers provide in-plane exchange fields to the superconductors.  
We obtain the spectrum of ABSs as a function of the Josephson phase in Fig.~\ref{figure1}(b).
The relative orientation of the in-plane exchange fields present in the superconductors splits the spin-degenerate ABSs. 
The degree of spin splitting of the ABSs is maximum for antiparallel alignment of the in-plane fields and increases with the strength of the exchange fields.
The ABSs diminish with increasing strength of the in-plane fields due to pair-breaking effect.
The behaviors of the ABSs are directly reflected in the CPR which follows a non-sinusoidal behavior [see the red curve in Fig.~\ref{figure1}(d)].
The quantum circuit theory~\cite{circuit99} is employed to numerically calculate the supercurrent for the system in Fig.~\ref{figure1}(a) for arbitrary transparencies. 
We then construct a different junction by adding a fully spin-polarized FM barrier between the superconductors as shown in Fig.~\ref{figure1}(c).
The total CPR is then a sum of the CPRs of the nonmagnetic and FM barriers. 
This results in a nonreciprocal CPR, or diode effect, as seen from Fig.~\ref{figure1}(d).
The diode efficiency is highly tunable by varying the relative orientation of the exchange fields and the conductance ratio between the FM and nonmagnetic barriers. 
Our system offers a practical approach that eliminates the need for flux piercing, unlike the Josephson diode effects studied earlier in Refs.~\cite{diode_Souto, fominov_JDE}.

{\it Andreev bound states.--}
An Ising superconductor in contact with an FM layer, with a singlet order parameter $\Delta$ and superconducting phase $\phi_\alpha$ can be described by a Bogoliubov–de Gennes Hamiltonian near the $\bm{K}$ ($\bm{K}'$) valley by neglecting the contribution from $\Gamma$ point~\cite{Kuz21}. 
The effective Hamiltonian can be written in the Nambu basis ($c_{\bm{k}, \uparrow}, c_{\bm{k}, \downarrow},c^{\dagger}_{-\bm{k}, \uparrow}, c^{\dagger}_{-\bm{k}, \downarrow}$) as
\begin{equation}
    H_{BdG} = \begin{bmatrix}
        H_0(\bm{k}) & \Delta e^{i\phi_\alpha} i \sigma_y \\
        -\Delta e^{-i\phi_\alpha} i \sigma_y & -H_0^*(-\bm{k})
    \end{bmatrix} ,
\end{equation}
where $H_0(\bm{k})$ is given by
\begin{equation}
    H_0 (\bm{k} = \bm{p} + s\bm{K}) = \xi_p \sigma_0 + s\beta_{so} \sigma_z - \bm{J} \cdot \bm{\sigma} .
\end{equation}
Here, $s = +1$ corresponds to the valley $\bm{K}$ and $s = -1$ to $\bm{K}'$. The deviation of the momentum from $\bm{K}$ or $\bm{K}'$ is denoted by $\bm{p}$.
The Pauli matrices $\sigma_x$, $\sigma_y$, and $\sigma_z$ act on the spin space, and $\sigma_0$ is the corresponding unit matrix.
The dispersion measured from the chemical potential $\mu$ is 
$\xi_p = |{\bm p}|^2 /2m - \mu$. 
The ISOC strength is denoted by $\beta_{so}$. The Zeeman term $\bm{J} \cdot \bm{\sigma}$ arises from the in-plane exchange field $\bm{J}$ provided by the FM layer. 
Note that due to the exchange field the main superconducting gap is reduced to
$2\Delta_{\rm eff}$ with $\Delta_{\text{eff}} = \Delta \beta_{so} / J_{\rm eff}$ and $J_{\rm eff} = \sqrt{\beta_{so}^2 + J^2}$.

We first consider a model based on the scattering approach for two identical Ising superconductors separated by a nonmagnetic scatterer in a single channel approximations [see Fig.~\ref{figure1}(a)]. 
The two superconductors have identical ISOC and we assume the relative angle between the two exchange fields provided by the two adjacent FM layers is denoted as $\theta$.
The condition for ABSs in the junction with ideal transmission can be found through determinant relation~\cite{Beenakker-ABS_1991}
\begin{equation} \label{det}
    \det(1 - R_1 S_1 R_2 S_2) = 0 \,.
\end{equation}
The Andreev scattering at each superconductor $\alpha$ ($= 1,2$) is described by a spin rotation matrix $R_\alpha$ and an electron-hole scattering matrix $S_\alpha$ given by
\begin{equation}
    R_{\alpha} = 
    \begin{bmatrix}
    e^{i \gamma \bm{n}_{\alpha} \cdot \bm{\sigma}} & 0 \\
    0 & e^{-i \gamma \bm{n}_{\alpha} \cdot \bm{\sigma}}
    \end{bmatrix}, \
    S_{\alpha} = \begin{bmatrix} 
    0 & r_e(\phi_{\alpha})\sigma_0 \\
    r_h(\phi_{\alpha})\sigma_0 & 0
    \end{bmatrix} .
\end{equation}
The unit vectors $\bm{n}_1$ and $\bm{n}_2$ denote the orientation of the in-plane exchange fields with $\bm{n}_1 = (1, 0, 0)$ and $\bm{n}_2 = (\cos\theta, \sin\theta, 0)$. 
The angle $\gamma = \arccos(\beta_{so} / J_{\rm eff})$ is between the $z$ axis and the effective field $\bm{J}_{\rm eff}$ as shown in Fig.~\ref{figure1}(a).
The Andreev reflection coefficients are $r_{e/h}(\phi_\alpha) = \exp(\pm i\phi_\alpha + i\chi)$ with superconducting phase $\phi_\alpha$ and
$\chi = \arccos(E /\Delta_{\text{eff}})$. 
Simplifying Eq.~\eqref{det} we obtain $\phi + 2\chi\pm \phi_\theta = 0$ where $\phi=\phi_1-\phi_2$ is the Josephson phase and $\phi_\theta$ is the spin splitting parameter defined 
\begin{equation}  \label{phi_theta}
    \cos \phi_\theta = \cos^2\gamma + \sin^2\gamma \cos\theta.
\end{equation}
This leads to the ABS energies given by
\begin{equation} \label{abs}
  E = \pm \Delta_{\rm eff} \cos\left[(\phi \pm \phi_\theta)/2\right] 
\end{equation}
and shown in Fig.~\ref{figure1}(b) as line.
The parameter $2\phi_\theta$ denotes the difference between the zero crossing of the ABSs and is due to the interplay between non-collinear exchange fields and ISOC. Obviously, from Eq.~\eqref{phi_theta} the spin-splitting parameter $\phi_\theta$ vanishes at $\theta=0$ or $J=0$. 
Since the function $\arccos(x)$ decreases monotonically in the range of $-1<x<1$, $\phi_\theta$ is maximal at $\theta=\pi$ and increases with increasing the exchange fields. 
The supercurrent carried by spin-split ABSs is given by
\begin{equation} \label{I_ana}
    I_0(\phi) = (2 e / \hbar) \sum (\partial E / \partial \phi)
    \tanh \big[E/(2k_BT) \big] ,
\end{equation}
where the summation runs over all the four solutions of $E$ in Eq.~\eqref{abs}. 

\begin{figure} 
    \centering
    \includegraphics[width=\columnwidth]{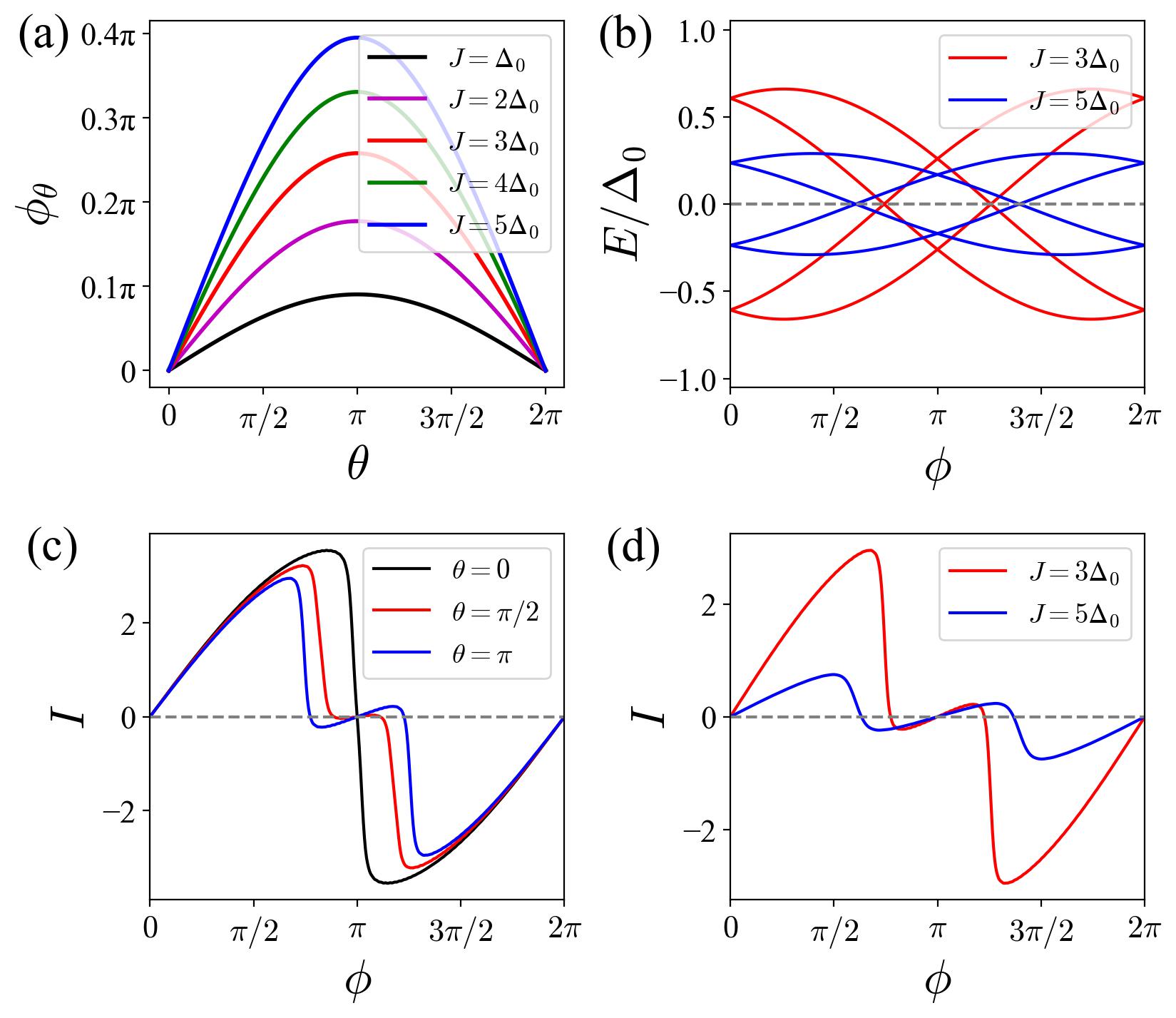}
    \caption{(a) The spin splitting parameter $\phi_\theta$ versus the relative magnetization angle $\theta$ at different exchange-field strengths $J$. (b) Andreev bound states at $\theta = \pi$. (c) Current-phase relations for various $\theta$ at $J=3\Delta_0$. (d) Current-phase relations at $\theta = \pi$. The supercurrent is in the units of $2G_0 \Delta_0 /e$.
    The temperature in (c) and (d) is $T=0.01 T_c$. }
    \label{figure2}
\end{figure}

To corroborate the simple picture, we perform calculations using quasiclassical Green's functions. In the following, we fix the  ISOC strength to $\beta_{so}=7\Delta_0$ where $\Delta_0$ is the superconducting gap in the absence of external fields at zero temperature. A broadening parameter of $0.01 \Delta_0$ is taken into account. The reduced superconducting order parameter $\Delta$ due to the in-plane exchange field and the finite temperature is calculated self-consistently. 
The ABSs given by Eq.~\eqref{abs} are shown in Fig.~\ref{figure1}(b) for $\theta=\pi/2$ and they overlap the numerically calculated spectral supercurrent obtained using Green's function methods discussed later.
The $4\pi$ periodic cosine curves in Fig.~\ref{figure1}(b), indicate spin-split ABS with the splitting is characterized by $\phi_\theta$.
Figure~\ref{figure2}(a) shows the dependence of $\phi_\theta$ on relative
magnetization angle $\theta$ for various exchange fields and it agrees with the
previous discussion on Eq.~\eqref{phi_theta}. 
The spin-splitting is maximal when the in-plane fields are antiparallel, i.e., $\theta=\pi$. 
Moreover, the splitting increases with the exchange field and reaches a maximum
at a critical value where the superconductivity is destroyed.
As the superconducting gap diminishes with increasing the exchange fields, the ABSs also diminish as seen from Fig.~\ref{figure2}(b). The ABSs eventually vanish at the critical value of the exchange fields.

The analytical treatment above provides results for a perfectly transparent nonmagnetic barrier. 
We now employ the quantum circuit theory~\cite{SM, circuit99} to obtain the CPRs at arbitrary transparencies of the junction. The expression of the supercurrent for a single-channel junction with transparency $D$ is given by 
\begin{equation} \label{ic}
  I (D,\varphi) = \frac{G_0}{8e}\int_{-\infty}^{\infty} d\varepsilon \ 
  {\rm tr}\left[ \tau_3\sigma_0\hat{I}^K(\varepsilon) \right], 
\end{equation}
with $G_0=2e^2/h$ being the conductance quantum. The Keldysh component of the matrix current $\hat{I}^K$ is given by
\begin{equation} \label{Ik}
  \hat{I}^K = 2 \left( \hat{A}^R \hat{X}^K + \hat{A}^K \hat{X}^A \right) ,
\end{equation}
with 
$ \check{A} = 2 D \big[\check{g}_{2}, \check{g}_{1}\big]$ and $ 
  \check{X} = \big[4 - D\big(2-\{\check{g}_{2}, \check{g}_{1}\}\big) \big]^{-1}$, where $\big[\check{g}_{2}, \check{g}_{1}\big]$ is the commutator and $\{\check{g}_{2}, \check{g}_{1}\}$ is the anti-commutator for the Green's functions. 
The Green's function $\check{g}_\alpha$ with $\alpha = 1, 2$ denoting the two superconductors is in Keldysh space and has the structure of
\begin{equation}
  \check{g}_\alpha = 
  \begin{pmatrix}
    \hat{g}^R_\alpha & \hat{g}^K_\alpha \\ 0 & \hat{g}^A_\alpha
  \end{pmatrix} ,
\end{equation}
where $\hat{g}^R_\alpha$, $\hat{g}^A_\alpha$, and $\hat{g}^K_\alpha$ are, respectively, the retarded, advanced, and Keldysh components of $\check{g}_\alpha(\varepsilon)$.
The calculation details for the Green's function are provided in the Supplemental Material~\cite{SM}.

The spectral supercurrent $I_{cs}$ defined as ${\rm tr}\big[\tau_3\sigma_0\hat{I}^K(\varepsilon) \big]$ for $D=1$ is plotted in Fig.~\ref{figure1}(b) and it agrees well with the ABSs given by Eq.~\eqref{abs}.
To understand how ABSs influence the supercurrent we compute CPRs using Eq.~\eqref{ic} as shown in Fig.~\ref{figure2}(c).
At the parallel magnetization configuration, we observe the well-known saw-tooth shape.
The supercurrent becomes increasingly non-sinusoidal as the magnetization configuration changes from parallel to antiparallel as seen in Fig.~\ref{figure2}(c).
This change directly reflects the increasing spin-splitting of ABSs.
For an antiparallel configuration as shown in Fig.~\ref{figure2}(c), the current switches parity and vanishes at two different phases other than $\phi = 0$ and $\phi=\pi$, due to a pair of ABSs crossing zero energy at these phases.
This signifies the presence of the second harmonics in addition to the first one in the CPRs.
Figure~\ref{figure2}(d) shows the decreasing supercurrent at stronger in-plane fields.
This is expected as the diminishing amplitude of the ABSs at higher fields [see Fig.~\ref{figure2}(b)] contributes to lower supercurrent.

{\it Superconducting diode effect.--}
A simple approach to obtain nonreciprocal critical currents for forward and reverse biases in a superconducting junction involves utilizing two different CPRs, as discussed in Ref.~\cite{fominov_JDE,diode_Souto}. 
Specifically, a junction with a CPR that is the sum of two different types: $I \propto \sin{(\phi + \phi_0)}$ and $I \propto \sin{(2\phi)}$ with nonvanishing $\phi_0$ is required for the diode effect. 
In our work, we have already achieved CPRs with components of $\sin(2\phi)$ using a transparent nonmagnetic barrier. 
Previous research~\cite{GT21} has demonstrated that phase-shifted sinusoidal CPR can be realized in an FM-Ising Josephson junction even in the tunneling limit. 
Based on these findings, we perform below a numerical assessment to achieve the diode effect.

As depicted in Fig.~\ref{figure1}(c), we combine a nonmagnetic junction having perfect transmission with an FM junction in the tunneling limit to form a Josephson diode. 
The nonmagnetic barrier contains $N_c$ transparent channels with conductance $G_N=N_c G_0$ so that the supercurrent is $I_N(\phi) = N_c I(1,\phi)$ obtained from Eq.~\eqref{ic}.
The FM channels are assumed to be fully spin-polarized with a magnetization pointing out of plane.
The supercurrent through the FM barrier is expressed as~\cite{SM, GT21}
\begin{equation} \label{I_F}
     I_{F} = I_{Fc} \sin{(\phi + \theta)}, 
\end{equation}
where $I_{Fc}$ is
\begin{equation}
    I_{Fc} = \frac{G_F}{8e}\int_{-\infty}^{\infty} d\varepsilon \; {\rm tr}\Big(\tau_3\sigma_0\hat{I}^R\Big) \tanh \big[\varepsilon / (2k_BT) \big] ,
\end{equation}
with the conductance of the FM tunnel barrier $G_F$. 
The retarded matrix current $\hat{I}^R$ is given by
\begin{equation}
    \hat{I}^R = {\rm Re} \Big[\hat{g}^R_2 + \{\hat{\kappa}, \hat{g}^R_2\} + \hat{\kappa}\hat{g}^R_2\hat{\kappa}, \ \hat{g}^R_1\Big] ,
\end{equation}
with the spin matrix $\hat{\kappa} ={\rm diag} (\sigma_z, \sigma_z)$.
The total current $I_{T}$ in this system would then be
\begin{equation} \label{I_T}
    I_T = I_N + I_F .
\end{equation}
It should be noted that the conductances $G_N$ and $G_F$ differ in the number of transmission channels and the transmission values. This implies that the conductance ratio $G_F/G_N$ in principle be controlled by the respective numbers of channels. 

The CPRs for $I_N$, $I_F$, and $I_T$ are plotted in Fig.~\ref{figure1}(d). The CPR for the nonmagnetic barrier with perfect transmission deviates from the sinusoidal behavior. The current vanishes at four values of $\phi$ including $\phi = 0$ and $\phi=\pi$. 
The CPR for the FM tunnel barrier is perfectly sinusoidal shifted by the relative magnetization angle $\theta$ of the two superconductors. 
For the total current of the system $I_T$, there exists a nonreciprocal critical current for forward and reverse bias directions.
Thus, a Josephson diode effect can be achieved. 
The diode efficiency $\eta$ that measures the nonreciprocity of the critical currents is defined as
\begin{equation}
    \eta = \frac{|I_{T}^+ - I_{T}^-|}{I_{T}^+ + I_{T}^-} , 
\end{equation}
where $I_{T}^+$ and $I_{T}^-$ are, respectively, the absolute values of the maximum and minimum of the total supercurrent $I_T$. 
For the case with conductance ratio $G_F / G_N = 10$ and relative exchange-field angle $\theta=0.9\pi$, the diode efficiency is as large as about $0.4$ in Fig.~\ref{figure1}(d). In the following, we provide a systematic investigation on the diode efficiency.

\begin{figure}
    \centering
    \includegraphics[width= \columnwidth]{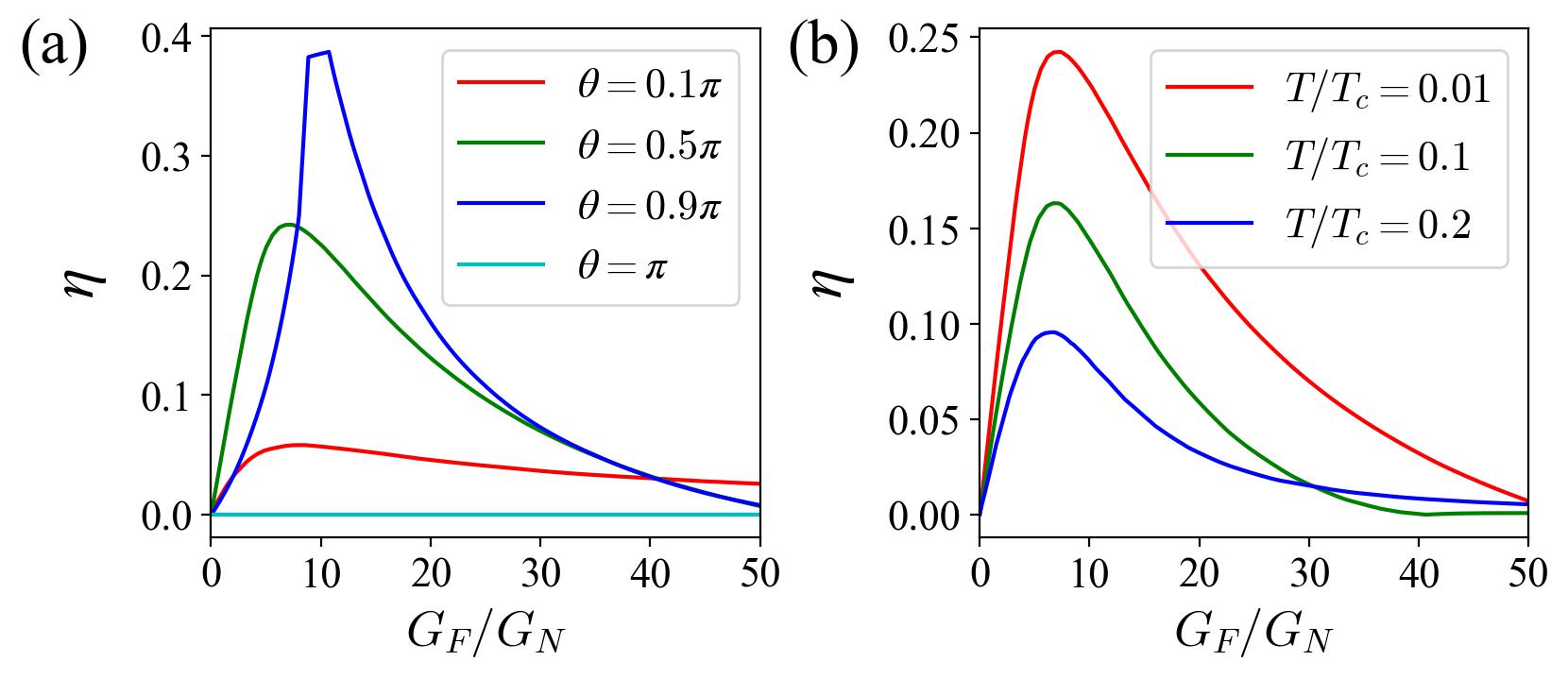}
    \caption{(a) Efficiency $\eta$ versus conductance ratio $G_F/G_N$ for different $\theta$ at $T = 0.01 T_c$. (b) Efficiency $\eta$ versus $G_F/G_N$ for different temperature $T$ at $\theta = \pi/2$.}
    \label{figure3}
\end{figure}

\begin{figure}
    \centering
    \includegraphics[width= \columnwidth]{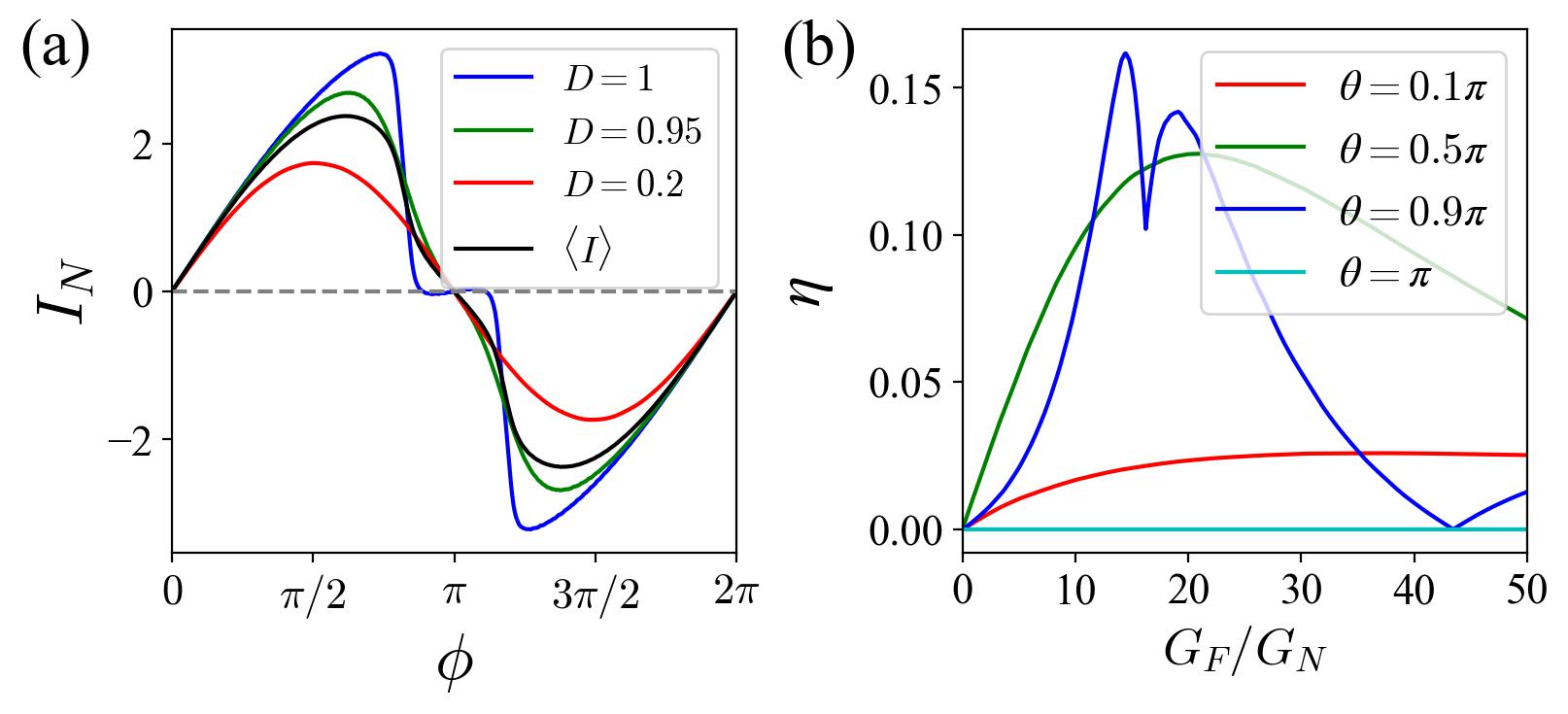}
    \caption{(a) Current-phase relations for different transparencies of the nonmagnetic barrier with $\theta=\pi/2$. The black curve represents the average current-phase relation using the Dorokhov distribution for a diffusive short junction. The supercurrents are in the units of $2G_N \Delta_0 /e$. 
    (b) Efficiency $\eta$ versus conductance ratio $G_F/G_N$ for different $\theta$ where the transmission in the nonmagnetic barrier follows the Dorokhov distribution. The temperature is $T=0.01T_c$.}
    \label{figure4}
\end{figure}

Figures~\ref{figure3}(a) and \ref{figure3}(b) display the diode efficiency as a function of the conductance ratio for various relative exchange-field angles and temperatures, respectively. The diode efficiency vanishes under the parallel and antiparallel configurations of the exchange fields where the CPR for the FM barrier follows $I_{Fc}\sin\phi$ or $-I_{Fc}\sin\phi$ as shown in Eq.~\eqref{I_F}. For a given conductance ratio, there is an optimal $\theta$ for the largest diode efficiency.
The diode efficiency is sensitive to temperature and decreases with it as seen from Fig.~\ref{figure3}(b). This is due to the fact that increasing temperature reduces the nonsinusoidal behavior or the second harmonics in the CPR of the nonmagnetic barrier. As shown in Fig.~\ref{figure4}(a), the CPR becomes more sinusoidal with decreasing transmission $D$ for the junction of nonmagnetic barrier.

In order to model a more realistic situation we now assume a large number of channels with a distribution of transmissions~\cite{Nazarov_QT}. As a generic example, we use the Dorokhov distribution of transmissions given by $\rho(D) = G_N/(\pi G_0 D \sqrt{1-D})$ with $G_N$ being the conductance of the diffusive part. The supercurrent for such a short diffusive nonmagnetic barrier is~\cite{Nazarov_QT}
\begin{equation}
    I_N(\phi) = \int_0^1 dD \rho(D) I(D, \phi) . 
\end{equation}
The corresponding CPR is displayed as the black curve in Fig.~\ref{figure4}(a). Remarkably, the diffusive CPR is still quite nonsinusodial, which means that the diode efficiency could be substantial.
We explore the nonreciprocity of the junction by combining the short diffusive nonmagnetic barrier and an FM tunneling barrier, where the total current is $I_T = I_N + I_F$. As shown in Fig.~\ref{figure4}(b) a significant diode efficiency is present in this more realistic disordered many-channel situation~\cite{diode_ando, diode_field-free}. This gives hope that our predicted diode effect is robust against disorder and might be accessible to experiments.

{\it Conclusion.--}
We have investigated the transport properties of Ising-superconductor Josephson junctions under the influence of in-plane exchange fields.
For a Josephson junction with a transparent nonmagnetic constriction, the ABSs are spin split due to the interplay of the exchange fields and ISOCs. 
The spin splitting can be controlled by the strengths and the relative orientation of the exchange fields. The properties of the ABS are directly reflected in the supercurrent.
We further demonstrate a Josephson diode effect by adding an FM tunneling barrier to the Ising-superconductor Josephson junction with the nonmagnetic channel of high transmission. 
The diode efficiency is tunable by varying either the relative exchange-field orientation or the conductance ratio of the barriers. Furthermore, the effect is insensitive to disorder and should be observable in generic Andreev-Ising-Josephson contacts with magnetic elements.
Our work provides a practical scheme in realizing the superconducting diode with two-dimensional heterostructures.
 
\begin{acknowledgments}
{\it Acknowledgements.--}
We acknowledge useful discussions with Michael Hein.
S.P. and W.B. acknowledge funding by the Deutsche Forschungsgemeinschaft (DFG, German Research Foundation)–Project-ID 443404566 - SPP 2244. G.T. is supported by National Natural Science Foundation of China (Grants No. 12088101 and 12374048) and NSAF (Grant No. U2330401).
\end{acknowledgments}

\bibliography{bib_IsingSC}{}
\clearpage

\begin{center}
  \large{\bf{Supplemental Material for ``The Andreev-Ising-Josephson Diode"}}
\end{center}

\section{The supercurrent from the quasiclassical Green's function}
The superconducting gap and the ISOC are much smaller compared to the Fermi energy, which allows us to use the formalism of quasiclassical Green's function~\cite{Eilenberger1968, LO1969,Belzig99,noneqSC, Eschrig15}. The structure of the Green's functions can be written as 
\begin{equation} \label{gg}
    \hat{g} (\hat{\bm{k}}, \varepsilon) = 
    \begin{bmatrix}
    g_0 \sigma_0 + \bm{g} \cdot \bm{\sigma}  & 
    (f_0 \sigma_0 + \bm{f} \cdot \bm{\sigma}) i \sigma_y \\
    (\Bar{f_0} \sigma_0 + \Bar{\bm{f}} \cdot \bm{\sigma}^*) i \sigma_y  & \Bar{g}_0 \sigma_0 + \Bar{\bm{g}} \cdot \bm{\sigma}^* 
    \end{bmatrix}.
\end{equation}
The bar operation in the above expression is defined as $\Bar{q} (\hat{\bm{k}}, \varepsilon) = q (-\hat{\bm{k}}, -\varepsilon^*)^*$ with $q \in \{ g_0, f_0, \bm{g},\bm{f} \} $. 
The retarded and advanced counterparts of $\hat{g}(\hat{\bm{k}},\varepsilon)$ are
obtained, respectively, by replacing the real $\varepsilon$ in $\hat{g}(\hat{\bm{k}},\varepsilon)$ with $\varepsilon + i\delta$ and $\varepsilon - i\delta$, where $\delta$ is an infinitesimal positive number.
The anomalous Green's functions $f_0$ and $\bm{f}$ describe the singlet and triplet pairings, respectively. 
By imposing the condition ${\rm tr}\big( \hat{g} \big) = 0$, we find that $\bar{g}_0 = -g_0$. 
Introducing ${\bm g}_{\pm}=({\bm g}\pm \bar{\bm g})/2$ ensures that the normalization condition $\hat{g} \hat{g} =\tau_0\sigma_0$ gives us 
\begin{align}
  &g_0^2 + \bm{g}_+^2 + {\bm g}_-^2 - f_0\bar{f}_0 +{\bm f}\cdot \bar{\bm f} =1,
  \label{norm} \\
  &2g_0 \bm{g}_+ = \bar{f}_0 {\bm f}- f_0 \bar{{\bm f}}, \label{g+} \\
  &2g_0 \bm{g}_- = i \bar{\bm f}\times {\bm f}. \label{g-}
\end{align}
We can write the Eilenberger equation for a homogeneous system in the clean limit as 
\begin{equation} \label{Eilen}
    [\varepsilon \sigma_0 \tau_3 - \hat{\Delta} - \hat{\nu} , \hat{g}]  = 0 ,
\end{equation}
combined with the conditions ${\rm tr}\big(\hat{g}\big)=0$ and the normalization $\hat{g}\hat{g}=\sigma_0\tau_0$. The order parameter term $\hat{\Delta}$ consisting of both the singlet and triplet components can be written as 
\begin{equation}
    \hat{\Delta} = 
    \begin{bmatrix}
    0 & \Delta e^{\phi_\alpha} i \sigma_y \\
    \Delta e^{-\phi_\alpha} i \sigma_y & 0
    \end{bmatrix}.
\end{equation}
The exchange and ISOC fields are included in the term $\hat{\nu}$ as
\begin{equation}
    \hat{\nu} = s \beta_{so}\sigma_z \tau_3 - 
        \begin{bmatrix}
            \bm{J} \cdot \bm{\sigma} & 0 \\
            0 &  \bm{J} \cdot \bm{\sigma^*} 
        \end{bmatrix}.
\end{equation}

Simplifying the above equations and combining them with Eq.~\eqref{g+}, which can be written as
\begin{equation}
  g_0 g_{+,x} = a\ \varepsilon J_x f_0 , \quad  g_0 g_{+,y} = a\ \varepsilon J_y f_0 , 
\end{equation}
we obtain $g_0$ and $f_0$ as
\begin{equation} \label{g0f0}
  g_0 = a\ \varepsilon \ c/(2\Delta), \quad 
  f_0 = -a\ (J^2 + c/2), 
\end{equation}
where
\begin{equation}
  c = \varepsilon^2-\beta_{\rm so}^2-J^2-\Delta^2 + u ,
\end{equation}
with 
\begin{equation} \label{u}
  u =  \sqrt{(\varepsilon^2-\beta_{\rm so}^2-J^2-\Delta^2)^2-4J^2\Delta^2} .
\end{equation}
The term $g_{-,z}$ is obtained from Eq.~\eqref{g-} with
\begin{equation}
  g_0 g_{-,z} = a^2\ \varepsilon J^2 \ s\beta_{\rm so} . 
\end{equation}
The coefficient $a$ is fixed by Eq.~\eqref{norm}, which is 
\begin{equation}
  g_0^2 +g_{+,x}^2 +g_{+,y}^2 +g_{-,z}^2 -f_0\bar{f}_0 +f_x\bar{f}_x +f_y\bar{f}_y =1 ,
\end{equation}
so that
\begin{equation}
  a^2 (4 J^2\Delta^2 -c^2)\big[ \Delta^2 (2 J^2 +c)^2 -c^2\varepsilon^2 +4 J^2
  \beta_{\rm so}^2 \Delta^2 \big] = 4 c^2 \Delta^2 ,
\end{equation}
which can be simplified as
\begin{equation}
  a^2 =\frac{c \Delta^2}{u^2 [ c(\varepsilon^2-\Delta^2)-2\Delta^2J^2 ]}.
\end{equation}
Having obtained the Green's functions in an Ising superconductor, we can proceed to calculate the supercurrent in the Ising-superconductor Josephson junctions.

Let us first consider a Josephson junction consisting of two Ising superconductors separated by a small nonmagnetic constriction.
The superconducting phase difference is denoted as $\phi$ and the relative angle between the two in-plane exchange fields as $\theta$.
From the quantum circuit theory~\cite{circuit99}, the expression of the supercurrent for a single-channel junction with transparency $D$ is given by 
\begin{align}
  I &= \frac{G_0}{8e}\int_{-\infty}^{\infty} d\varepsilon \ 
  {\rm tr}\big(\tau_3\sigma_0\hat{I}^K \big),
\end{align}
with $G_0=2e^2/h$ the conductance quantum.
In the above, $\hat{I}^K$ is the Keldysh component of the matrix current given by
\begin{equation}
  \hat{I}^K = 2 \big( \hat{A}^R \hat{X}^K + \hat{A}^K \hat{X}^A \big) ,
\end{equation}
with
\begin{equation}
  \check{A} = 2 D \big[\check{g}_{2}, \check{g}_{1}\big] , \quad 
  \check{X} = \Big[4 - D\Big(2-\big\{\check{g}_{2}, \check{g}_{1}\big\}\Big)\Big]^{-1} .
\end{equation}
We have included an additional factor of $2$ in Eq.~\eqref{Ik} to take the valley degrees of freedom into account. 
Here, $\check{g}_\alpha$ with $\alpha = 1, 2$ is in the Keldysh space and has the structure 
\begin{equation}
  \check{g}_\alpha = 
  \begin{pmatrix}
    \hat{g}^R_\alpha & \hat{g}^K_\alpha \\ 0 & \hat{g}^A_\alpha
  \end{pmatrix} ,
\end{equation}
where $\hat{g}^R_\alpha$ and $\hat{g}^A_\alpha$ are, respectively, the retarded and advanced counterparts of $\check{g}_\alpha$. The Keldysh component can be obtained via the relation
\begin{equation}
  \hat{g}^K_\alpha = \tanh\big[\varepsilon/(2k_B T)\big] \big( \hat{g}^R_\alpha - \hat{g}^A_\alpha \big)
\end{equation}
with the Boltzmann constant $k_B$ and the temperature $T$. 

For an Ising-superconductor Josephson junction with a ferromagnetic barrier, the supercurrent is given by~\cite{GT21}
\begin{align}  \label{IF}
  I_F &= I_{Fc} \sin{(\phi + \theta)},
\end{align}
with
\begin{equation}
    I_{Fc} = \frac{G_F}{8e}\int_{-\infty}^{\infty} d\varepsilon \; {\rm tr}\Big(\tau_3\sigma_0\hat{I}^R \Big) \tanh \big[\varepsilon / (2k_BT) \big],
\end{equation}
where $G_F$ is the conductance of the ferromagnetic tunnel barrier and the retarded matrix current is
\begin{equation}
    \hat{I}^R = {\rm Re} \Big[\hat{g}^R_{2} + \{\hat{\kappa}, \hat{g}^R_{2}\} + \hat{\kappa}\hat{g}^R_{2}\hat{\kappa}, \ \hat{g}^R_{1}\Big] ,
\end{equation}
with the spin matrix $\hat{\kappa} ={\rm diag}({\bm m}\cdot {\bm \sigma}, {\bm m}\cdot {\bm \sigma}^*)$. 
The valley degrees of freedom have been taken into account in Eq.~\eqref{IF}.
We consider the case where the magnetization of the barrier points out of plane
with ${\bm m}=(0,0,1)$. By considering a clean limit of the Ising superconductors, we have
\begin{align} \label{kgg}
  & {\rm Re}\big[{\rm tr}\big(\tau_3 \sigma_0\big[ \{\hat{\kappa}, \hat{g}_{2}^R\}, 
    \hat{g}_{1}^R\big]\big)\big] \notag \\
  =& -16\, {\rm Im}(a^2) (\varepsilon^2 + \beta_{\rm so}^2) J^2 \sin\theta \cos\phi 
  + \cdots ,
\end{align}
and 
\begin{align} 
  & {\rm Re}\big[{\rm tr}\big(\tau_3\sigma_0\big[ \hat{\kappa}\hat{g}_{2}^R\hat{\kappa},
  \hat{g}_{1}^R\big]\big)\big] \notag \\
  =& 8\, {\rm Im}\big[ f_{0}\bar{f}_{0} - a^2 (\varepsilon^2 + \beta_{\rm so}^2)
  J^2 \cos\theta \big] \sin\phi ,
\end{align}
where the terms which are odd in valley index are neglected in Eq.~\eqref{kgg}.

\end{document}